\documentclass[aps,prb,showpacs,floatfix,superscriptaddress,twocolumn]{revtex4-1}

\pdfoutput=1

\usepackage{natbib} 
\usepackage{amsmath} 
\usepackage{amssymb} 
\usepackage{graphicx}

\usepackage{fancyref}

\def\etal{\mbox{et al.}}

\begin{document}

\title{On the reaction coordinate for seeded crystallisation} 

\author{Swetlana Jungblut}
\affiliation{Faculty of Physics, University of Vienna, Boltzmanngasse 5, 1090 Wien, Austria}
\author{Christoph Dellago}
\affiliation{Faculty of Physics, University of Vienna, Boltzmanngasse 5, 1090 Wien, Austria}

\begin{abstract}

Small pre-structured seeds introduced into an undercooled fluid are known to increase the
crystal nucleation rate by some orders of magnitude, if the structure of the seeds is commensurate
with the bulk crystalline phase. The presence of such seeds also alters the crystallisation mechanism by favouring
particular structures at the early stages of the nucleation process. Here, we study with computer simulations the effect of small
face-centred cubic and body-centred cubic seeds  on the crystallisation of a Lennard-Jones liquid in the strongly undercooled regime. We find that seeds with
body-centred cubic structure lead to a larger enhancement of the crystallisation rate than face-centred
cubic seeds. An analysis of recurrence times reveals that the size of the largest crystalline cluster used as
reaction coordinate is affected by pronounced memory effects, which depend on the particular seed structure
and point to the importance of structural information in the definition of a good reaction coordinate for crystallisation.

\bigskip

\bigskip

\end{abstract}

\maketitle 


\section{Introduction}\label{intro}

The crystallisation of undercooled liquids is a widely studied topic in recent research with many technological implications \cite{gasser:2009, vekilov:2010, herlach:2010, anwar:2011, tanaka:2012, sear:2012, christenson:2013,palberg:2014, sear:2014}. Due to its fundamental and practical importance, this process has been studied extensively in simple models, which can be realized experimentally in colloidal suspensions. A qualitative picture of the crystallisation process is provided by classical nucleation theory, which posits that a small crystalline nucleus forms within the undercooled liquid. Due to the creation of a crystal-liquid interface surrounding the nucleus, this process is free-energetically uphill in its initial stages. Driven by thermal fluctuations, the crystalline nucleus may nevertheless grow until it reaches a critical size after which further growth is thermodynamically favourable eventually leading to the crystallisation of the entire system. While the basic perspective provided by classical nucleation theory is essentially correct, its details are still subject of current discussions. Several computer simulation studies performed for hard sphere, Gaussian core, or 
Lennard-Jones (LJ) systems \cite{omalley:2003, schilling:2010, karayiannis:2011, lechner:2011, russo:2012, russo:2012a, tenwolde:1996,moroni:2005, beckham:2011,jungblut:2013a} agree that the structure of the crystalline clusters formed in the course of the transition is not uniform and reorganizes as the reaction proceeds. In LJ crystallisation \cite{tenwolde:1996,moroni:2005, beckham:2011,jungblut:2013a}, studied also in this article, the body-centred cubic (bcc) structure is formed first and subsequently transforms into the face-centred cubic (fcc) structure, such that the crystalline clusters have, on average, an fcc-structured core and a bcc-structured surface. While fcc is the thermodynamically stable phase in bulk LJ crystals and the bcc phase is only metastable, the initial formation of bcc crystals is favoured kinetically by a lower free energy barrier, thus providing an instance of Ostwald's step rule \cite{ostwald:1897, stranski:1933}. This empirical rule says that a metastable system may transform into its final state through formation of an intermediate phase if the free energy barrier between the initial and the intermediate (bcc) phases is lower than the one between the initial and the thermodynamically most stable (fcc) states.

For a detailed description of crystallisation and the computation of free energies, a reaction coordinate quantifying the progress of the transition is needed. For this purpose, the number of particles in the largest crystalline cluster, 
identified using the scheme proposed by ten Wolde, Ruiz-Montero, and Frenkel \cite{tenwolde:1996} 
based on the Steinhardt bond order parameters \cite{steinhardt:1983}, has been widely used.  Although recognized to have its faults, this reaction coordinate still performs best in comparison with other suggested collective variables \cite{beckham:2011, jungblut:2013a}. The deficiency of the size of the largest crystalline cluster becomes particularly apparent \cite{jungblut:2015} in the calculation of crystal nucleation rates within the framework of mean first-passage time (MFPT) analysis \cite{kramers:1940, haenggi:1990, bartell:2006, wedekind:2007, kashchiev:2007}. In this approach, which is practical only for large undercooling, one carries out straightforward molecular dynamics simulations starting in the metastable liquid. From the mean time required to first reach a certain size of the largest crystalline cluster one can then extract the nucleation rate as well as the size of the critical nucleus. The MFPT method relies on the assumption that the chosen reaction coordinate evolves diffusively according to the Smoluchowski equation \cite{risken:1989,gardiner:2009}. In particular, when analysing mean first passage times to extract reaction rate constants, one implicitly surmises that the time evolution of the reaction coordinate is Markovian, i.e. the future of the reaction coordinate depends only on its current but not its past values. The assumption of Markovianity also implies that the dynamics of the reaction coordinate does not depend on any other variable of the system. As shown recently \cite{jungblut:2015}, however, these assumptions are not valid for the crystallisation transition described using the size of the largest crystalline cluster as reaction coordinate. Indeed, an analysis of recurrence times reveals that the dynamics of this reaction coordinate is strongly non-Markovian, leading to significant errors in the crystallisation rates estimated from mean first passage times \cite{jungblut:2015}. This non-Markovianity observed for a LJ system is due to the neglection of other degrees of freedom that are important for the transition, such as the structure of the crystalline nucleus. Interestingly, the violation of the assumptions underlying the MFPT analysis is not evident in the mean first passage times themselves, explaining why this behaviour has not been noticed earlier. 

Here, we further study the effects of a poor choice of reaction coordinate, using crystallisation on pre-structured seeds as example. By introducing small seeds with various structures into the undercooled liquid, one can modify the crystallisation mechanism by favouring the nucleation of particular arrangements and inhibiting the formation of others. Most of the previous studies on seeded crystallisation \cite{cacciuto:2004, devilleneuve:2005, browning:2008, dullens:2008, schoepe:2011, hermes:2011, vanteeffelen:2008, neuhaus:2013} considered either unstructured and rather large seeds \cite{cacciuto:2004, devilleneuve:2005, browning:2008, dullens:2008, schoepe:2011} or two-dimensional systems \cite{vanteeffelen:2008, neuhaus:2013}. In three dimensions, Hermes \etal ~\cite{hermes:2011} compared experimental results on crystallisation on small two-dimensional structured templates with computer simulations. In this work, we use comparatively small three-dimensional seed structures, the size of which is just large enough to make the distinction between the fcc and bcc structures possible. In previous studies \cite{jungblut:2011a, jungblut:2013}, we have shown that the effect of the seed is related to its structure in the sense that the commensurability with the bulk equilibrium structure is one of the factors which influence the crystallisation rate. In particular, we found that the seeds with a regular fcc structure produced the largest increase of the reaction rate. This tendency is expected, but the increase of the reaction rate by several orders of magnitude was rather surprising considering the size of the seeds. In this paper, we study seeded crystallisation at a slightly larger undercooling and find that, in this case, the largest increase of the crystallisation rate is obtained with bcc rather than fcc seeds. This result is rather counterintuitive, because in the bulk the bcc structure is only metastable. As in the case of homogeneous crystallisation \cite{jungblut:2015}, the analysis of recurrence times reveals that the size of the largest crystalline nucleus evolves in a non-Markovian way, pointing to the necessity to include additional collective variables such as structure and shape into the description of the nucleation mechanism. 

The remainder of the article is organized as follows. We start with a description of the simulation details and then present reaction rates obtained with transition interface path sampling (TIS) \cite{vanerp:2003} and MFPT techniques for four types of the seeds with fcc and bcc structures and different values of the lattice spacing. These results are complemented with an analysis of recurrence times and of critical nuclei determined based on committor calculations. We conclude the paper with a discussion focused on the ability of the largest crystalline cluster to describe the crystallisation mechanism and on artifacts arising due to a poor choice of the reaction coordinate.

\section{Simulations}
\label{model}

We perform molecular dynamics (MD) simulations in the $NpH$ ensemble \cite{andersen:1980}, which permits to accommodate the density change occurring during the crystallisation and, at the same time, avoids the artificial removal of latent heat by a thermostat.  The integration time step is $\Delta t = 0.01$ (in LJ units, which are used throughout the paper). The pressure is set to a value close to zero ($p=0.003257$) and the enthalpy per particle $H/N=-5.11$ is chosen such that the initial temperature is $T=0.5$, corresponding to approximately $28\%$ undercooling \cite{agrawal:1995}. The particles, evolving in a cubic simulation box with periodic boundary conditions, interact via the truncated and shifted LJ potential with a cutoff distance of $r_c=2.5$. We simulate systems with seeds of different structure. The seed particles consist of either $13$ (fcc) or $15$ (bcc) fixed particles arranged in a way around the central particle to reproduce the first shell of the corresponding structure. We study the effect of seed particles with the lattice spacing of the bulk crystal, $d=1.09$, and squeezed seeds with a smaller lattice spacing of $d=1.0$. For seeds with $d=1.0$ and $d=1.09$, the number of freely moving particles is $N=6636$ and $N=6627$, respectively. The crystalline clusters are identified by the standard scheme \cite{tenwolde:1996} based on the Steinhardt bond order parameters \cite{steinhardt:1983}. If considered without the surrounding particles, the cluster analysis recognizes only the central particle of the seed as solid.  

We compute crystal nucleation rates by means of transition interface sampling simulations (TIS) \cite{vanerp:2003} using the number of particles in the largest crystalline cluster, $n_s$, to define the interfaces. The TIS interfaces are positioned at $n_s=30, 50, 80, 120, 170, 230, 300, 400$ for seeds with $d=1.0$ and $n_s=30, 70, 130, 230, 400$ for seeds with $d=1.09$. These TIS simulations yield the conditional probability that the crystalline nucleus reaches a given size after having left the initial undercooled state defined by $n_s\leq 20$. This probability initially decreases as a function of $n_s$, but then reaches a plateau for cluster sizes $n_s$ beyond the free energy barrier that separates the metastable liquid state from the fully crystalline phase. The crystallisation rate is obtained as the product of the plateau value of the conditional probability to reach the crystalline state and the flux out of the initial undercooled liquid state. In the original version of TIS \cite{vanerp:2003}, the flux out of the initial state is calculated in a straightforward MD simulation of the undercooled liquid by counting the number of times the system crosses the first (TIS) interface ($n_s=30$) when coming directly from the initial state. This approach relies on the presence of a sufficiently high free energy barrier between the states, such that the system spends most of the simulation time in the initial state (meanwhile, the TIS method has been modified also in this respect \cite{vanerp:2007,bolhuis:2008}). This is, however, not the case for the conditions we consider here, since the system is strongly undercooled and the low free energy barrier does not prevent crystallisation on the timescale of the simulations. In addition, the presence of the seeds flattens the free energy landscape in the vicinity of the initial phase and the system, even when not crystallising, spends a considerable fraction of time out of the initial state. 

It may appear that, under the strong undercooling conditions studied here, direct methods such as the MFPT approach \cite{kramers:1940, haenggi:1990, bartell:2006, wedekind:2007, kashchiev:2007} provide a more convenient way for the calculation of nucleation rates. However, as found recently \cite{jungblut:2015}, the MFPT analysis is affected by the non-Markovianity of the time evolution of the largest cluster size used as reaction coordinate resulting in large inaccuracies in the rate estimate. We will discuss this issue in detail in the next section. The memory effects leading to non-Markovian behaviour, however, are localized close to the transition region such that the MFPT analysis can be still used to calculate the flux out of the initial state required in TIS. We estimate this flux as the inverse of the MFPT at the first TIS interface for $200$ crystallising trajectories starting in the initial state ($n_s\leq 20$).  

To identify configurations containing critical clusters, we performed a committor analysis \cite{dellago:2002} on configurations collected from crystallising paths obtained in TIS simulations. For every kind of seed, we selected about $1500$ configurations with various cluster sizes $n_s$. For each configuration, we started $100$ trajectories with momenta picked randomly from a Maxwell-Boltzmann distribution and let the system evolve in time until it reached either the liquid ($n_S=20$) or fully crystalline ($n_s=1000$) state. The commitment probability, $p_B$, of a particular configuration is then given by the fraction of crystallising trajectories. The transition state ensemble consists of configurations with $p_B\approx 0.5$. 

\section{Results and discussion}\label{results}

In this section, we compare the effects of four different seed structures on the crystallisation rates and critical cluster sizes. In addition, we present the reaction rates obtained with the MFPT technique finding that the deviations from the TIS results are much larger than in the case of bulk crystallisation at the same conditions. We also examine the memory effects that lie at the origin of the inaccuracies of the MFPT rate calculations. 

\subsection{Crystal nucleation rates}

\begin{figure}[tb]
\begin{center}
\includegraphics[clip=,width=0.95\columnwidth]{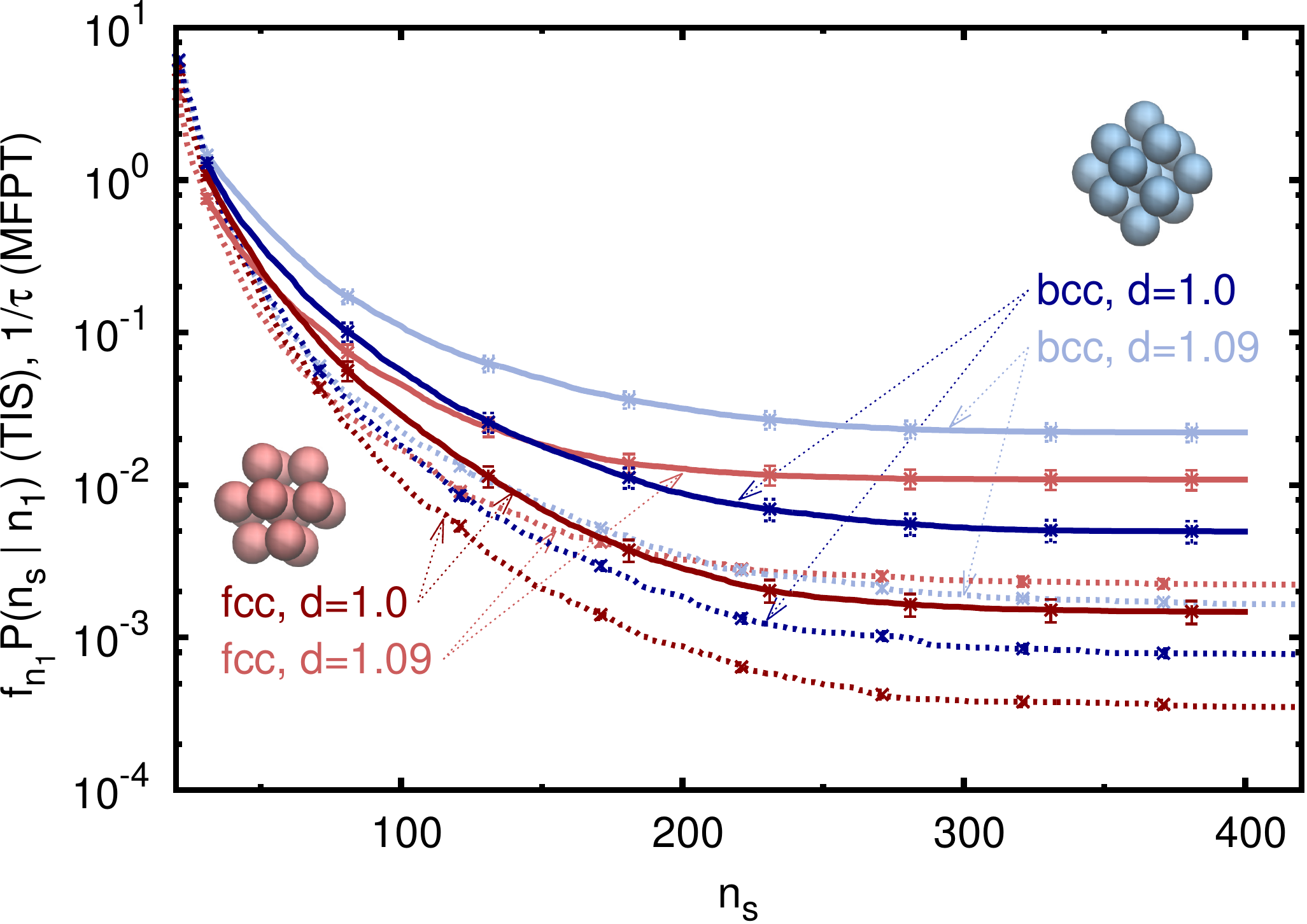} 
\caption{\label{rates} Product of the flux out of the initial state with the conditional probability to reach a given cluster size $n_s$, $f_{n_1}P(n_s|n_o)$ (solid lines), and inverse MPFTs, $1/\tau$ (broken lines), for systems with bcc($d=1.0$), bcc($d=1.09$), fcc($d=1.0$), and fcc($d=1.09$) seeds (as indicated). Both quantities saturate at the value of the crystallisation rate, $JV$. For clarity, statistical errors are indicated only for selected data points.} 
\end{center}
\end{figure}

We start with the crystallisation rates obtained from the TIS simulations.  In Fig.~\ref{rates}, we present the product of the flux $f_{n_1}$ through the first TIS interface with the probability $P(n_s|n_o)$ to reach a given 
crystalline cluster size $n_s$ under the condition that the system initially started in the undercooled liquid state. For larger clusters, when the system has crossed the free energy barrier between the initial and final states of the transition, the conditional probability $P(n_s|n_o)$ becomes constant and leads to the crystallisation rate: 
\begin{equation}\label{tisrate}
JV = f_{n_1}P(n_s=400|n_o), 
\end{equation}
where $V$ is the volume of the simulation box, $J$ is the nucleation rate (number of nucleation events per unit time and unit volume), $n_1$ is the position of the first TIS interface, and $n_0=20$ 
is the boundary of the initial state. As mentioned in the previous section, we use the inverse of the MFPT at the first TIS interface to  compute the flux out of the undercooled state. 

In contrast to the results obtained previously for a smaller degree of undercooling ($25\%$) \cite{jungblut:2011a, jungblut:2013}, the presence of the bcc-structured seed with bulk lattice spacing induces the largest increase of 
the crystallisation rates. The squeezed seeds cause a comparatively smaller increase of the reaction rate, but also here, the bcc-structured seed raises the  rate more than the fcc-structured one. In terms of Ostwald's step rule \cite{ostwald:1897, stranski:1933}, which relates the transition path to the height of the free energy barrier the system has to overcome to leave the metastable state, we can connect the behaviour of our system to the free energy landscape underlying the reaction. The presence of both seeds lowers the free energy barrier to the corresponding state, but then the bcc structure still has to transform to the fcc phase. For the smaller undercooling \cite{jungblut:2011a, jungblut:2013}, we found that the heights of the free energy barriers to the bcc and fcc phases, lowered due to the presence of seeds, are comparable, since the probabilities to reach smaller cluster sizes are similar for both systems. As the cluster size increases, the probability in the presence of an fcc seed becomes constant, while the probability in the system with a bcc seed continues to decrease, resulting in a lower crystallisation rate. This is not the case at the higher undercooling studied here. For all sizes considered, regular as well as squeezed bcc seeds lead to larger probabilities to reach a given cluster size than the corresponding fcc seeds. Thus, the barrier to the bcc phase in the presence of a bcc seed has to be lower than the barrier to the fcc phase with an fcc seed. In addition, the difference in the barrier heights appears to be larger than the height of the free energy barrier between the bcc and fcc phases. 

Figure~\ref{rates} contains also the inverse MFPTs, $1/\tau$, measured in straightforward MD simulations in the presence of the corresponding seeds. It is evident from the figure that the crystallisation rates obtained from the MFPT analysis are considerably smaller than the respective rates determined in the TIS simulations. As discussed previously \cite{jungblut:2015}, this difference originates from the inability of the size $n_s$ of the largest cluster to capture the essential features of the crystallisation mechanism. This poor choice of reaction coordinate results in the non-Markovian character of the process and distorts the MFPT analysis. 
The deviations between the MFPT and TIS rates are even larger than in the case of bulk crystallisation at the same conditions, where rates differ by a factor of two \cite{jungblut:2015}. Here, the deviation depends on the type of the seed and varies between a factor of four (for fcc seeds) and a factor of over 13 (for bcc seeds), as specified in Table~\ref{ratestab}. 

These findings indicate that a good reaction coordinate for the crystallisation transition has to include information not only about the size of the crystalline clusters but also about their structure. Structural effects are even more pronounced for seeded crystallisation than for bulk crystallisation, because the pre-structured seeds modify the structural composition of small crystalline clusters, thus, shifting the preferred path along the free energy landscape in this structural contribution to the reaction coordinate. 

 \begin{table}
 \caption{\label{ratestab} Reaction rates $JV$, calculated as $f_{n_1}P(n_s=400|n_o)$ (TIS) and 
$1/\tau(n_s=400)$ (MFPT), their ratio for the systems with four types of the seeds. Also included are values computed for bulk crystallisation \cite{jungblut:2015} Note that, for the particular value of $n_s=400$ used for the rate calculations, the probability $P(n_s=400|n_o)$ has reached a plateau, as can be inferred from Fig.~\ref{rates}, indicating that the system has crossed the nucleation barrier. }
 \begin{center}
 \begin{tabular}{l|c|c|c}
   &TIS  $\times 10^{-2}$& MFPT $\times 10^{-3}$ &Ratio\\
   \hline
   \hline
   bcc, $d=1.0$& $0.50 \pm 0.08 $ &$0.78 \pm 0.06$&$6.33516$\\
   bcc, $d=1.09$&$2.22 \pm 0.30$&$1.66 \pm 0.12$&$13.3548$\\
   fcc, $d=1.0$&$0.15 \pm 0.03$&$0.35  \pm 0.03 $&$4.19111$\\
   fcc, $d=1.09$&$1.09 \pm 0.16 $&$2.23 \pm 0.16 $&$4.87689$ \\
   bulk &$0.017 \pm 0.004 $&$0.083 \pm 0.006 $&$1.99543$ 

 \end{tabular}
 \end{center}
 \end{table}

\subsection{Recurrence times}
\begin{figure}[tb]
\begin{center}
\includegraphics[clip=,width=0.95\columnwidth]{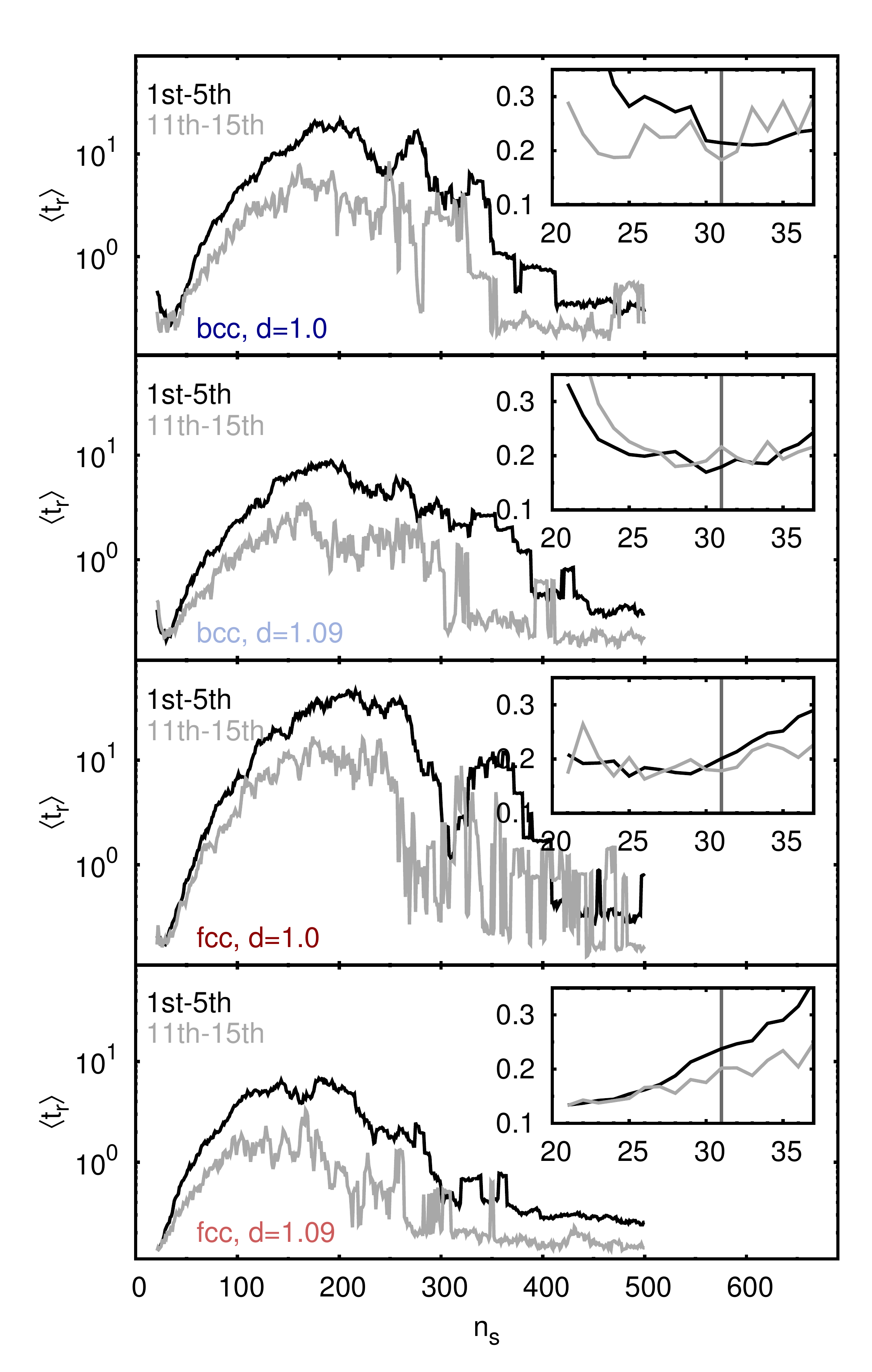} 
\caption{\label{rtimes} Mean recurrence times as a function of the largest cluster size, $n_s$, for the systems with (as indicated) bcc ($d=1.0$), bcc ($d=1.09$), fcc ($d=1.0$), and fcc ($d=1.09$) seeds. To improve the statistics, we averaged the times over five crossings, i.e. ten recurrences, in each case: $1st$--$5th$ (black lines) and $11th$--$15th$ (grey lines). Insets: Close-up view on the corresponding data in the area around the first TIS interface (vertical line).} 
\end{center}
\end{figure}
The non-Markovian character of the crystallisation described in terms of the size of the largest crystalline cluster, $n_s$, is most clearly seen in the behaviour of the recurrence times as discussed in the following.
For a given cluster size, we consider not only the first passage through an imaginary interface corresponding to this size, as in the MFPT approach, but also subsequent passages through this interface in the same direction. The recurrence time is then defined as half of the period between two successive crossings. In a Markov process, the recurrence times, which are inversely proportional to the stationary probabilities of states \cite{smoluchowski:1915, kac:1947}, should not depend on the number times the interface has been crossed before. In fact, for Markovian dynamics there is no memory of the past and so the average time between the first and the second crossing of the interface should be the same as the average time interval between any subsequent consecutive crossings of the interface. The data presented in Fig.~\ref{rtimes} clearly contradict this assumption. Evidently, the recurrence times averaged (to improve the statistics) over the first five crossings, i.e. ten recurrences, are distinctly larger (by up to a factor of about ten to hundred, depending on the type of the seed) than the times for later passages demonstrating that the dynamics of $n_s$ is indeed strongly non-Markovian.
\begin{figure}[tb]
\begin{center}
\includegraphics[clip=,width=0.95\columnwidth]{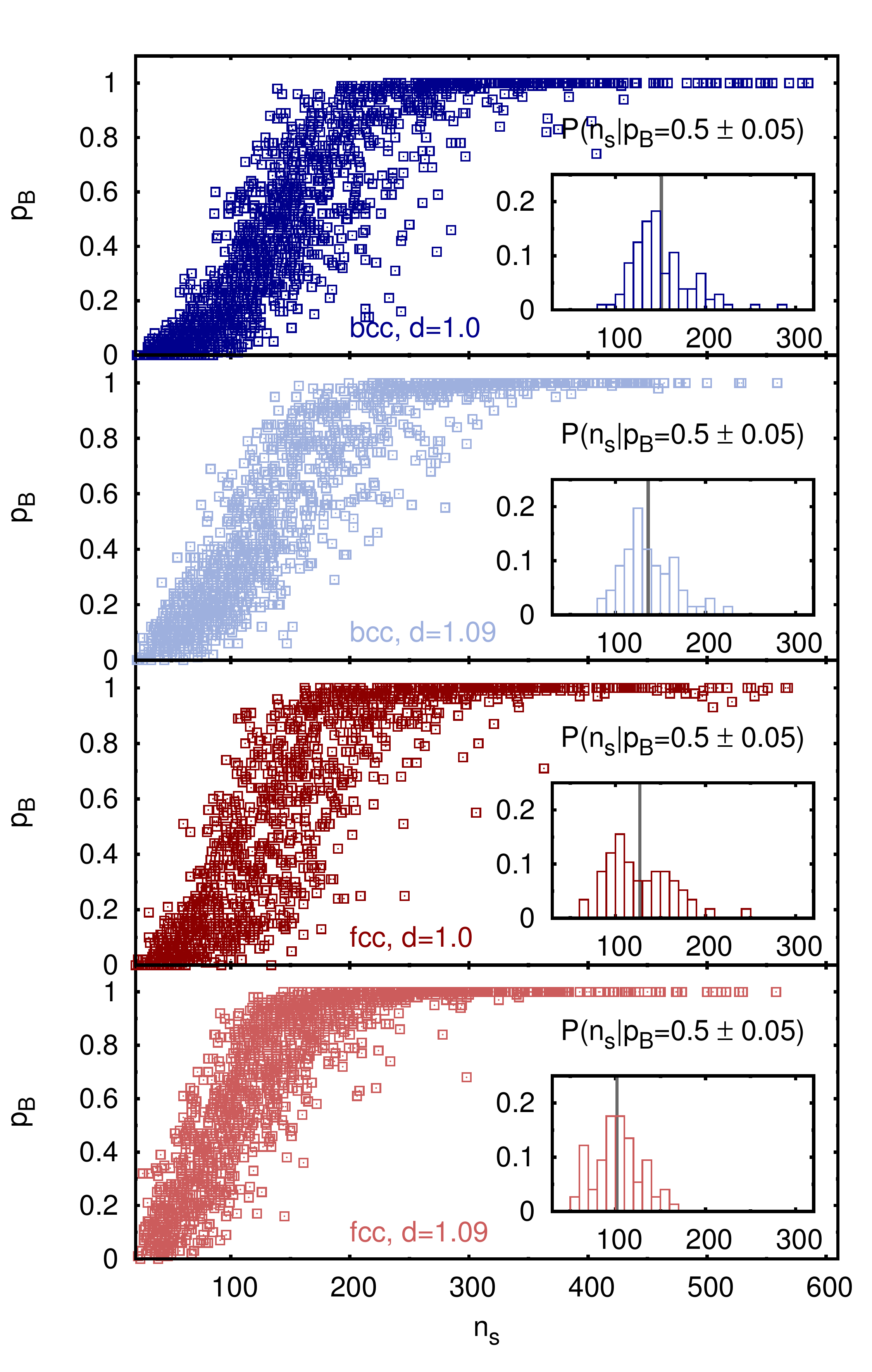} 
\caption{\label{commprobs} Committor, $p_B$, as a function of the largest cluster size, $n_s$, for the systems with  bcc ($d=1.0$), bcc ($d=1.09$), fcc ($d=1.0$), and fcc ($d=1.09$) seeds (as indicated). Insets: Corresponding size distributions of the critical clusters, identified from configurations that have a committor of $p_B=0.5\pm0.05$. The average critical cluster sizes are indicated by the vertical lines.} 
\end{center}
\end{figure}

While these memory effects are most pronounced in the transition region, in the presence of fcc seeds the non-Markovianity becomes noticeable even in the vicinity of the first TIS interface, as can be seen in the insets of Fig.~\ref{rtimes}. The fluctuations are rather large, but this behaviour is consistent with the observation that the values of the flux through the first TIS interface calculated in straightforward MD simulations (not presented here) deviate from the inverse MFPTs at this interface in the presence of the fcc seeds. In the MD simulations, we used rather short trajectories, ensuring that the system does not crystallise, and corrected for longer excursions out of the initial phase by considering for flux calculations not the total simulation time but only periods the system spent below the first TIS interface. The fluxes calculated in this way reproduce the inverse MFPTs in the presence of bcc seeds, but are slightly larger for fcc seeds. The effect is small but this finding is in accordance with the behaviour 
of the recurrence times --- in a longer MD simulation, nothing changes for bcc seeds, but the system 
crosses the first interface more often as the time proceeds in the presence of fcc seeds.

\subsection{Critical clusters}

We identify critical clusters within the commitment analysis as those found in configurations with equal probability to relax into the liquid state or to crystallise completely. Together, all these configurations form the transition state ensemble. The results of the commitment analysis of the systems with various seeds are presented in Fig.~\ref{commprobs}. As expected, the distribution of cluster sizes $n_s$ in the transition state ensemble is rather broad, indicating, once again, an insufficient reaction coordinate. However, we note a rather peculiar connection between the critical cluster size and the crystallisation rate: the regular bcc seed ($d=1.09$) induces the largest increase in the crystallisation rate but the critical cluster size is rather large. The critical clusters are, on average, largest for the squeezed bcc seed, and smallest for the regular fcc seed, which is, however, not reflected in the behaviour of the reaction rates. 

\begin{figure}[tb]
\begin{center}
\includegraphics[clip=,width=0.95\columnwidth]{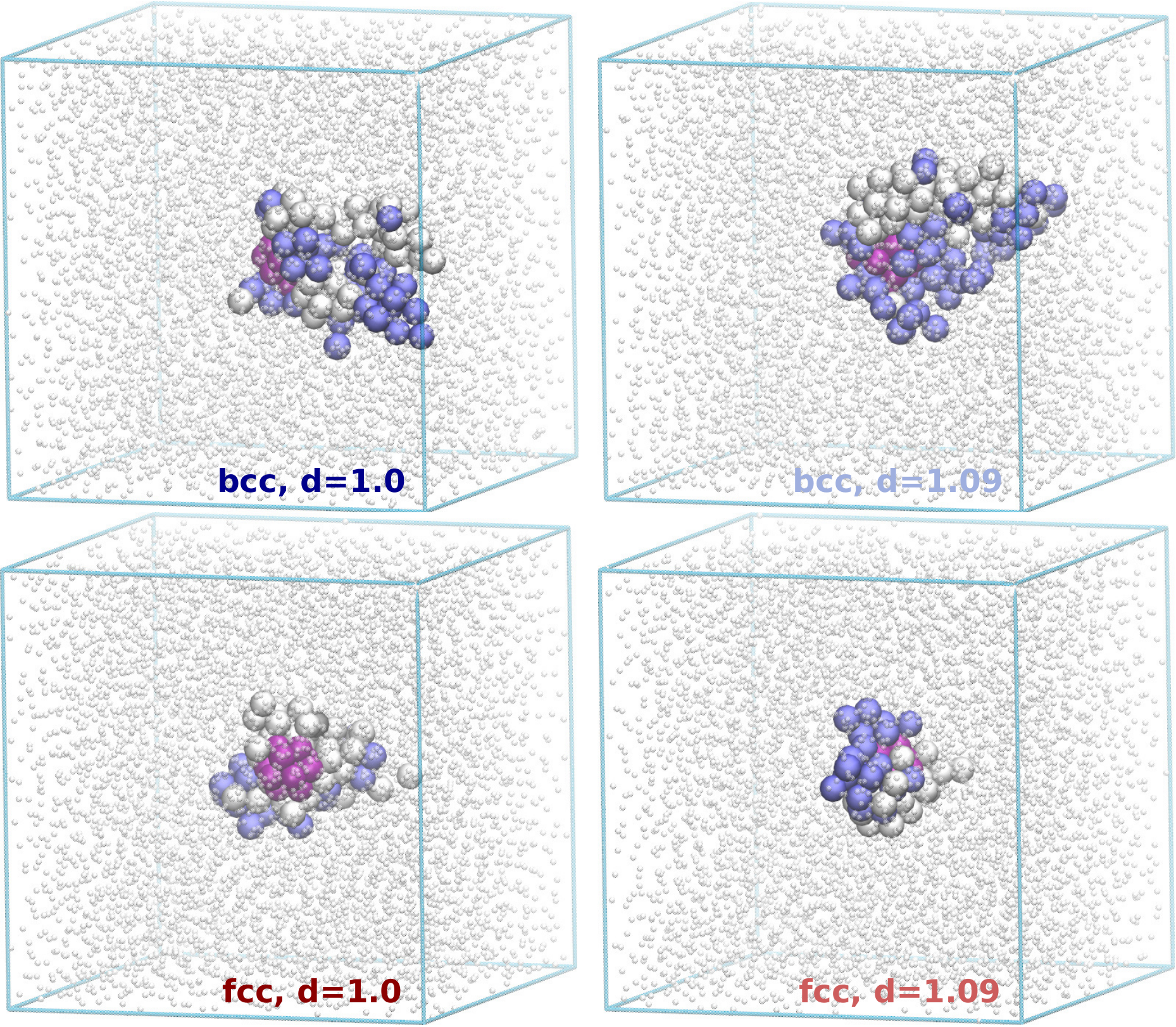} 
\caption{\label{snapshotsColor} Snapshots of the configurations containing critical clusters for bcc ($d=1.0$), bcc ($d=1.09$), fcc ($d=1.0$), and fcc ($d=1.09$) seeds (as indicated). 
Seed particles are depicted as purple spheres. Particles which are not part of the largest crystalline cluster are represented by small white spheres. Large white spheres indicate particles in the fcc/hcp environment, and blue spheres picture particles in the bcc structure.} 
\end{center}
\end{figure}

In Fig.~\ref{snapshotsColor}, we show some representative configurations belonging to the transition state ensemble. The critical clusters formed on the fcc-structured seeds have a mostly fcc-structured core and bcc-structured surface, although the size of the crystal does not allow a strict distinction between the surface and the core. Bcc seeds show the opposite tendency, displaying a bcc core and an fcc surface layer. 

For a closer look at the structures, we used a combination of locally averaged Steinhardt bond order 
parameters $w_4$ and $w_6$ \cite{lechner:2008, jungblut:2011} to identify various structures in the crystalline clusters. In doing so, 
we find not only perfect fcc and bcc structures, but also hexagonal close-packing (hcp) and 
a crystalline structure, newly discovered to be metastable in LJ systems \cite{eshet:2008}, which are distortions of the fcc and bcc structures, respectively. 
However, for clarity reasons, we refer to these distorted structures also as to fcc and bcc.         
  
\section{Conclusion}
\label{sum}

In this work, we have studied the influence of small seeds on the early stage of crystal nucleation in an undercooled LJ liquid paying particular attention to the quality of the nucleus size as reaction coordinate. The presence of the seeds with pre-defined structure modifies the transition pathways, thus allowing to accentuate and investigate the importance of structural aspects of the transition mechanism. 

In our TIS simulations, carried out in the regime of large undercooling, we find that bcc-structured seeds have the strongest effect on the crystallisation rate yielding an enhancement factor of over $130$ with respect to the bulk nucleation rate under the same conditions. This result is particularly surprising as the fcc structure and not the bcc structure is the thermodynamically preferred bulk phase of LJ crystals. In contrast, for a smaller degree of undercooling studied earlier \cite{jungblut:2013}, largest crystallisation rate was observed in the presence of an fcc seed with lattice spacing commensurate to the bulk crystal. This finding, which does not rely on a good definition of the reaction coordinate, indicates that the free energy landscape changes non-trivially with undercooling. 

The importance of structural properties for the crystallisation transition is also reflected in a pronounced non-Markovianity that arises in the dynamics of the largest crystalline cluster used to quantify the progress of the reaction. Our analysis of recurrence times, which are found to display striking memory effects, indicates that the projection of the reaction kinetics on this coordinate distorts the description of the crystallisation mechanism. A symptom of this problem is that the mean first passage time analysis, which is based on the assumption of Markovianity of the underlying dynamics, yields an incorrect estimate of the crystallisation rate. While the deviating reaction rates, the non-stationary recurrence times, and the counter-intuitive transition states found here indicate that important degrees of freedom have been neglected in the description of the transition mechanism, they do not directly point to which information needs to be included in the definition of a good reaction coordinate for crystallisation. In future research, a systematic variation of the seed properties combined with an analysis of the crystalline nuclei after a varying number of recurrences may lead to a progress on this important issue. 

\section*{Acknowledgments}

The computational results presented have been achieved using 
the Vienna Scientific Cluster (VSC). We acknowledge financial support of the 
Austrian Science Fund (FWF) within the Project V 305-N27 as well as SFB ViCoM (Grant F41). 

\bibstyle{revtex}

\end{document}